# Wide Separation Planets In Time (WISPIT): Discovery of a Gap Hα Protoplanet WISPIT 2b with MagAO-X

Laird M. Close*[1], Richelle F. van Capelleveen[2], Gabriel Weible[1], Kevin Wagner[1], Sebastiaan Y. Haffert[2,1], Jared R. Males[1], Ilya Ilyin[3], Matthew A. Kenworthy[2], Jialin Li[1], Joseph D. Long[4], Steve Ertel[1,5], Christian Ginski[6], Alycia J. Weinberger[7], Kate Follette[8], Joshua Liberman[9], Katie Twitchell[9], Parker Johnson[1], Jay Kueny[9], Daniel Apai[1], Rene Doyon[10], Warren Foster[1], Victor Gasho[1], Kyle Van Gorkom[1], Olivier Guyon[1,11,9,12], Maggie Y. Kautz[1], Avalon McLeod[13], Eden McEwen[9], Logan Pearce[14], Lauren Schatz[15], Alexander D. Hedglen[16], Ya-Lin Wu[17], Jacob Isbell[5], Jenny Power[5], Jared Carlson[5], Emmeline Close[18], Elena Tonucci[2], Matthijs Mars[2]

## ABSTRACT

Excellent (<25 mas) Hα images of the star TYC 5709-354-1 led to the discovery of a rare Hα protoplanet. This star was discovered by the WISPIT survey to have a large multi-ring transitional disk, and is hereafter WISPIT 2. Our Hα images of 2025, April 13 and April 16 discovered an accreting (Hα in emission) protoplanet: WISPIT 2b (r=309.43±1.56mas; (~54au deprojected) , PA=242.21±0.41°) likely clearing a dust-free gap between the two brightest dust rings in the transitional disk. Our SNR=12.5 detection gave an Hα ASDI contrast of $(6.5\pm0.5)\times10^{-4}$ and a Hα line flux of $(1.29\pm0.28)\times10^{-15}$ erg/s/cm$^2$. We also present L' photometry from LBT/LMIRcam of the planet (L'=15.30±0.05 mag) which, when coupled with an age of $5.1^{+2.4}_{-1.3}$ Myr, yields a planet mass estimate of 5.3±1.0 $M_{jup}$ from the DUSTY evolutionary models. WISPIT 2b is accreting at $2.25^{+3.75}_{-0.17}\times10^{-12}$ $M_{sun}$/yr. WISPIT 2b is very similar to the other Hα protoplanets in terms of mass, age, flux, and accretion rate. The inclination of the system ($i$=44°) is also, surprisingly, very similar to the other known Hα protoplanet systems which all cluster from 37≤$i$≤52°. We argue this clustering has only a ~1.0% (2.6σ) probability of occurring randomly, and so we speculate that magnetospherical accretion might have a preferred inclination range (~37-52°) for the direct (cloud free, low extinction) line of sight to the Hα line formation/shock region. We also find at 110mas (~15au deprojected) a close companion candidate (CC1) which may be consistent with an inner dusty 9±4 $M_{jup}$ planet.

---

* Corresponding author lclose@arizona.edu
Rest of the author affiliations at end of paper

## 1. INTRODUCTION

It is now well established that some gas giant protoplanets pass through a period of high luminosity as they accrete hydrogen gas from their circumplanetary disks producing detectable Hα emission. This was most clearly demonstrated initially in the discovery of Hα emission from PDS 70 b (K. Wagner et al. 2018), and PDS 70 c (S. Haffert et al. 2019). Direct observations of protoplanets (defined here as accreting planets) are a key window into this very poorly understood process of planet formation and accretion from a circumplanetary disk (CPD) which itself is embedded in a larger circumstellar disk, or transitional disk; Espaillat et al. 2011). While the exact mechanisms of planetary accretion are not yet fully understood, massive planets could magnetospherically accrete, via magnetic fields, directly onto a latitude line of the planet (Z. Zhu et al. 2016; T. Thanathibodee et al. 2019; Marleau et al. 2022 and references within). Accretion through shocks onto the circumplanetary disk is also possible (Y. Aoyama et al. 2018; J. Szulágyi & Mordasini 2017; Y. Aoyama et al. 2021 and references within), and it is unclear which process, or a combination of both, dominate. To be clear the Aoyama model and magnetospheric accretion model are not mutually exclusive. The Aoyama model could also explain the Hα emission in a magnetospheric accretion scenario. The difference is the origin of the emission. The Thanathibodee model assumes emission from the accretion flow tracing the magnetic field. The Aoyama model assumes emission from the shock itself. Variability studies may be able to inform which of these models are more likely (Demars et al. 2023; L. Close et al. 2025; Y. Zhou et al. 2025 and references within). The key to informing our accreting protoplanet models is to discover more systems –because there is only one really well studied system (PDS 70) to date (L. Close et al. 2025 and references within). Indeed, the study of protoplanets is critical if we are to understand the process of planet formation, accretion, satellite/moon growth, CPDs and the impact that these planets have on their host disks (clearing gaps, and creating cavities, etc.).

In section 2 of this letter, we briefly introduce the current state of Hα protoplanet detections, instrumentation and techniques. We introduce the newly discovered transitional disk star WISPIT 2 at the end of that section. Note, our companion letter, Letter 1, (R. van Capelleveen et al. 2025; hereafter Letter 1) covers the H+Ks characterization and discovery of the star's impressive multi-ringed transitional disk and planet in the NIR. In section 3, we describe our MagAO-X and LBTI/LMIRcam observations of WISPIT2. In section 4, we introduce the discovery Hα images of the protoplanet WISPIT 2b, and follow-up images at L'. Section 5 presents the Hα and L' photometry and astrometry of WISPIT 2b. In section 6, we analyze the Hα photometry to derive the line flux and mass accretion rate of WISPIT 2b. In section 7, we derive a mass for WISPIT 2b from the L' photometry, and compare to the H+Ks planetary mass from Letter 1. We also discuss WISPIT 2b compared to the other known Hα protoplanets (defined as exoplanets that have SNR>5 Hα emission detections at multiple epochs). At the end of this discussion section we describe an inner Close Companion (CC1) which could be an inner planet or an unusually red compact dust clump. Our conclusions are given in section 8.

## 2.0 MagAO-X INSTRUMENTAL CONFIGURATION FOR Hα IMAGING

### 2.1. Introduction to Hα Protoplanet Imaging

It is not trivial to detect protoplanets. The only way to guarantee an actively accreting protoplanet is being detected is to directly detect accretion tracers. Using the MagAO (the predecessor AO system to MagAO-X) system, L. Close et al. (2014) used the strongest visible tracer of accretion (Hα) to detect the low mass companion HD 142527 B inside the large transitional disk dust-free gap of HD 142527 A. The work of L. Close et al. (2014) first speculated that for low mass ($0.5<M_p<10$ $M_{jup}$) planets, Hα angular spectral differential imaging (ASDI) could be a powerful tool for detection of protoplanets, particularly at the lower mass end where Hα could be brighter than the NIR emission for active accretion. Indeed, using MagAO's SDI+ mode (L. Close et al. 2018) we discovered Hα emission from the PDS 70 b protoplanet (M. Keppler et al. 2018) in May of 2018 (Wagner et al. 2018). Then, VLT/MUSE discovered PDS 70 c at Hα (S. Haffert et al. 2019). Recently L. Close et al. (2025) has utilized the much improved MagAO-X SDI mode to capture PDS 70 b and c over 3 years showing great sensitivity to variable Hα emission from protoplanets.

### 2.2. New Hα detection Techniques with Extreme Visible AO: MagAO-X

Past "Hα AO" detections were executed with older AO systems (e.g. VLT/SPHERE, VLT/MUSE, Magellan/MagAO) with relatively low (<1-10%) Strehls at Hα. However, we have now fully commissioned the world's newest extreme AO system MagAO-X. MagAO-X is unique –it was designed from the start to work in the visible at high Strehl (J. Males et al. 2018; 2024). The optical design for MagAO-X is complex in that, being a woofer-tweeter system, requires 2 reimaged pupils, and a lower coronagraphic bench which requires another pre-apodizer pupil followed by a Lyot pupil plane. Hence MagAO-X has 4 reimaged pupils created by 8 off-axis parabolas (OAPs), this complex optical train could lead to thermal alignment drift and variable non-common path (NCP) errors –but each OAP is potted in our (L. Close & M. Kautz) patented ultrastable "set-and-forget" mounts (US Patent Number 11,846,828) which minimizes NCP thermal drift. MagAO-X has an 6x6" FOV at f/69 with 0.00593±0.00003"/pix platescale (with 13 micron EMCCD pixels) that yields a nicely oversampled 3.4 pix/($\lambda$/D) at H$\alpha$. See L. Close et al. (2018; 2025) for more detail about the optical design of MagAO-X.

MagAO-X yields a superior level of wavefront control with a 2040 actuator Tweeter deformable mirror (DM) and a unique "extra" DM to eliminate all NCP errors between the science and wavefront sensing channels, minimizing coronagraphic leak (we call this DM the NCPC DM). This NCPC DM was upgraded to 1024 actuators in 2024 which greatly improved our ability to use Focal Diversity Phase Retrieval (FDPR; K. Van Gorkom et al. 2021; J. Kuney et al. 2024) to eliminate NCP errors by an artificial source NCP closed-loop calibration/elimination at the start of the night (or after a major beamsplitter change during the night). This approach is our field-tested optimal procedure to minimize NCP errors and achieve uniquely high H$\alpha$ AO Strehls on faint targets.

On this run we FDPR calibrated away NCP errors with the NCPC DM to enable ~90-94% Strehl (no atmospheric turbulence) at H$\alpha$ on our science cameras with our "always ready" artificial source (a super-continuum laser) before the start of science operations and/or after a major beamsplitter change. It is typically only done once a night, and only takes ≤20 minutes usually after sunset in twilight with dome fully open near ( ±2C) our observing temperature. After which the artificial source is removed from the beam and the NCPC DM stays in this calibrated DM shape for the rest of H$\alpha$ imaging night and so the f/69 focus and NCPC wavefront is fixed –and no more observing time is lost to any other calibrations all night long.

Wavefront sensing (WFS) with MagAO-X's very low noise (<0.6 rms e- read noise) EMCCD pyramid WFS OCAM2 detector allows Strehls of >60% to be obtained at z' (908 nm; $\Delta\lambda$=130 nm) while closed loop at 2kHz (residual WFE <120nm rms) with 1564 corrected modes– as demonstrated on-sky (J. Males et al. 2022; 2024). The low noise of this WFS sensor enables good correction at H$\alpha$ even on faint I~10 mag guide stars in median ~0.65" seeing conditions (whereas, a laser guide star WFS system could not deliver better correction at Hα due to the cone effect; N. Siegler et al. 2007). The MagAO-X system with up to 1564 corrected modes maps to ~14 cm/actuator, making it the highest sampled AO system in the world. So deeper, much more sensitive, surveys for Hα planets are finally possible.

### 2.3. Introduction to TYC 5709-354-1 (WISPIT 2)

TYC 5709-354-1 is a 1.1 $M_{sun}$ classical T-Tauri star of age $5.1^{+2.4}_{-1.3}$ Myr (Letter 1) which is actively accreting. It was discovered by the WISPIT survey and VLT/SPHERE H band imaging to have a spectacularly large cavity with two bright outer rings and a dark gap centered at ~68 au followed by another bright ring (Letter 1). At a nearby distance of just 133pc (GAIA; L. Lindegren et al. 2021; and discussion in Letter 1) with its complex ring system and its young age TYC 5709-354-1 (henceforth WISPIT 2) was an excellent target for MagAO-X Hα SDI imaging to determine if there were any accreting protoplanets in the central cavity or annular gaps.

## 3.0 MagAO-X OBSERVATIONS OF WISPIT 2

### 3.1. The 2025 April 13 Observations of WISPIT 2

MagAO-X has a high throughput Hα SDI mode with all custom λ/10 beamsplitters (with ~95% transmission of Hα) where the Hα photons are transmitted to the two science cameras and only ~5% are lost to the wavefront sensor optical path (see L. Close et al. (2025) for details). Moreover, this mode also allows a very efficient SDI camera setup where another custom λ/10 beamsplitter cube transmits ~95% of the Hα continuum to a continuum filter ($\lambda_{CONT}$=668.0 nm; $\Delta\lambda_{CONT}$=8.0 nm) in science camera 1. This cube simultaneously reflects ~95% of the Hα light to narrowband Hα filter ($\lambda_{H\alpha}$=656.3 nm; $\Delta\lambda_{H\alpha}$ =1.045 nm) to science camera 2. For the optical design of these two EMCCD science cameras, see the left hand side of Fig. 1 in L. Close et al. (2025).

For clarity and completeness, we list all the environmental, instrumental, and reduction settings in Table A1 in Appendix A for each night WISPIT 2 was observed at Hα. Table A1 is also a summary of all the reduction settings/values for all our WISPIT 2 Hα observations.

On our first night (2025, April 13 UT) in slightly worse than average seeing (0.68-1.08"), we observed WISPIT 2 for two hours before transit. We were able to lock the AO loop on WISPIT 2 (I=9.9 mag) with 1000 AO modes at 1000 Hz with the gain of each mode set by MagAO-X's autogain feature. The best 64.35% of those 0.5s integration images at Hα had a FWHM of 25 mas (Strehl=8-12%) and led to a 1.23 hr final integration time.

MagAO-X is particularly well suited to the discovery of Hα protoplanets due to its "photon-counting" EMCCDs. Here we set the Hα camera to near its maximum gain, so that EMgain$_{Hα}$ = 294.13±0.29 ADU/e- (the effective readnoise=0.05e- rms; "photon counting" mode) in the Hα images. For the rest of the observing and instrumental set-up please see Table A1.

*3.2. The April 16, 2025 Observations of WISPIT 2*

Our 2$^{nd}$ night of observing WISPIT 2 was better than the first night. The seeing was excellent (0.34-0.52") and we obtained 56 degrees of rotation (25% more than the first night). A major change was to use the 50/50 science beamsplitter cube to simultaneously observe in the z' (908 nm, BW=131 nm) broad-band continuum filter (instead of the 688 nm, BW=8 nm Hα continuum filter). The science camera 2 remained in the Hα (656.3 nm, BW=1.045 nm) filter, as on the previous night, but throughput was now 50% of the maximum due to the 50/50 science cube. The AO correction was continuously excellent, we selected 99.5% of the Hα data (all of which had FWHM<24 mas) and so we had 2.16 hours of total integration (8,084x 1s frames at Hα and 32,336x 0.25s frames at z'). We should note that due to the selection of the 50/50 science beamsplitter and the Hα/IR WFS beamsplitter, the z' throughput was only ~25% maximum for z' –regardless the images were excellent and the very high Strehl at z' made up for any throughput losses. The final Hα PSF had a FWHM=23.6 mas resolution in the 2.16 hour image (Strehl~30%).

4.0 REDUCTIONS

Data reduction was with a custom pipeline (L. Close et al. 2025) which was designed around the fact that the flux from protoplanets at Hα is very low. Indeed, for PDS 70 b (which is similar to WISPIT 2b's Hα flux) we typically received only ~1 Hα planet photon/pixel every 20s –so only one in forty 0.5s images actually detects a single Hα photon on a given planet pixel (L. Close et al. 2025). This implies that one needs to average 120x 0.5s exposures together before there is a good chance of ~3 detected planet Hα photons per pixel within one FWHM sized patch (~4x4 pix) centered on the planet core. So binning in time (averaging) 120x 0.5s images to a 60s average image yields ~48 Hα photons/min from the planet's core spread over 16 pixels. This yields enough signal to noise ratio (SNR) for the planet signal to survive our high-pass filters and KLIP

PCA PSF fitting and removal procedures on these 60s images. Therefore, our custom python/pyIRAF pipeline (described in L. Close et al. 2025) was optimized for the preservation of individual photon events (each photon counted by the EMCCDs) while also maximizing the contrast with ADI and SDI (which together we call, hereafter, ASDI).

*4.1. Discovery of WISPIT 2b*

As described in Close et al. (2025), the pipeline averaged the selected frames into "averaged" frames of 60s. Then, as is usual for PCA PSF removal (J. Wang et al. 2015; Follette et al. 2023), these "averaged" frames were "pre-processed" to remove the low-spatial frequencies (the smooth "seeing halo") around each PSF. To do this we high-pass filtered the images by smoothing with a FWHM=57.3 mas Gaussian (purposely much larger than the ≤25mas FWHM of a planet, to preserve planet core flux) and subtracting this smoothed image from the original to remove the low-spatial frequencies, while preserving the point-sources (such as planets). Unwanted flux was further removed from PSF by fitting a radial profile (centered on the core) of the images and removed from each of these frames (again this step leaves the planet core untouched). Then all 74x 60s images (first night) and 130x 60s (second night) images were fed into pyKLIP (J. Wang et al. 2015) to remove the PSF via PCA in the usual manner in reducing ADI data. Then the last step is the accurate scaling of our non-coronagraphic data (multiplying the continuum by StarFlux$_{H\alpha}$ / StarFlux$_{CONT}$) and subtraction of this scaled continuum image from the H$\alpha$ image to yield the final ASDI image. The ASDI image should only trace true H$\alpha$ emission as all residual contamination from the PSF, or scattered light off dust, should have been subtracted off (see section 5 for more details). While we have in the past (Follette et al. 2023; Li et al. 2025) magnified the pixel scale of the continuum by 656.3/$\lambda_{CONT}$ to exactly match the H$\alpha$ diffraction rings, we skip this step here. Because we cannot change the spatial scale of the continuum as that shifts the positions of strong continuum emission from the dust rings, which would then poorly subtract (and actually exacerbate artefacts) at r~200-400 mas region of interest for these ASDI images. With WISPIT 2 we are limited more by dust structures (which rotate so pyKLIP cannot remove) than diffraction rings (which pyKLIP can mostly remove).

Once this process had been run on the 2025, April 13 dataset we discovered a SNR=5.5 H$\alpha$ emission point source near the middle of the outer annular gap of WISPIT 2 (see Fig. 1 top row). This exciting discovery motivated the 2025, April 16 dataset which recovered the planet

WISPIT 2b at SNR=12.5 at Hα (see Fig. 1 bottom row). There was no significant detection of continuum light (668nm) at the location of the very cool/red planet, marking it as a very rare example of a protoplanet with Hα emission –and the first one in an annular gap. We note that the PDS 70 planets are considered to be in the large central dust-free cavity (S. Haffert et al. 2019), while WISPIT 2b is outside of the central cavity, in an annular "ring" gap –sandwiched between two bright narrow dust rings, which is unique for an Hα protoplanet.

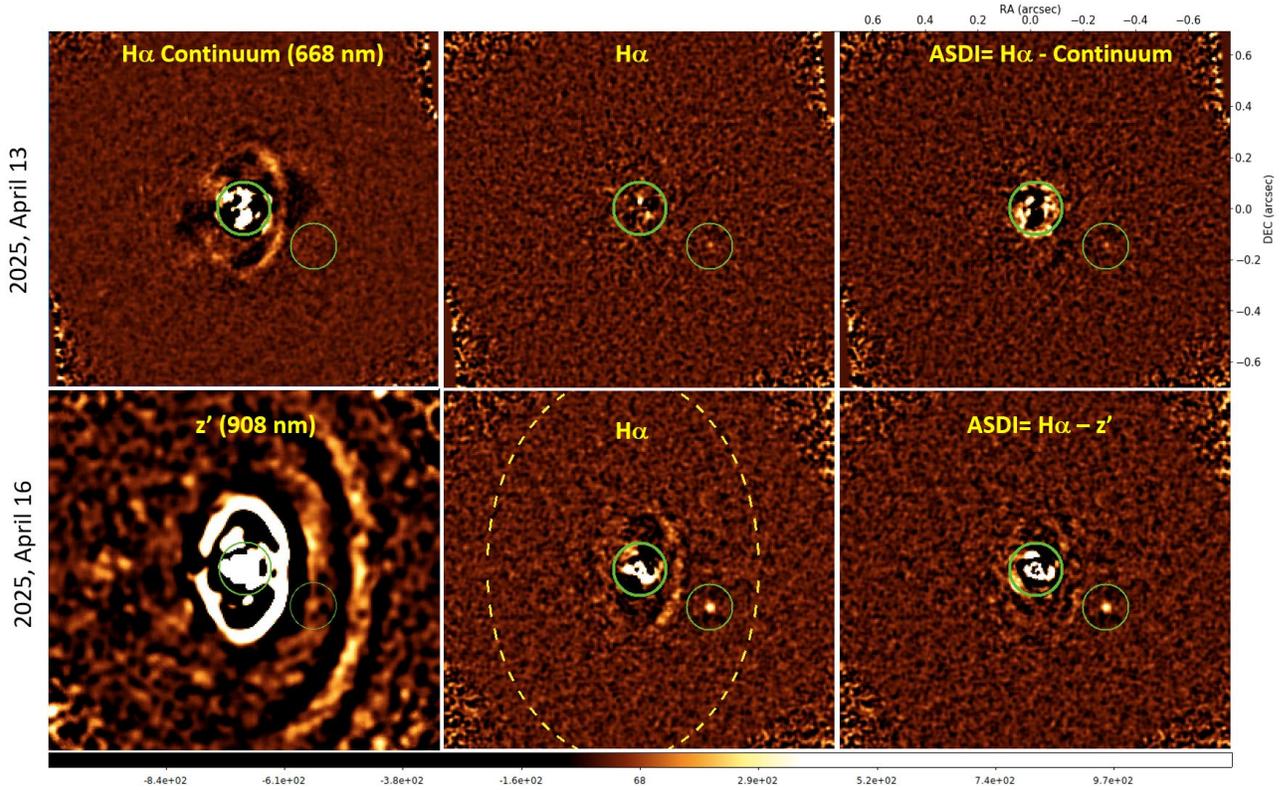

**Fig 1**: The discovery images of WISPIT 2b. These are both the April 2025 MagAO-X Hα pyKLIP pipeline reduced datasets. The thick central green circles (r=103 mas) are centered on the star. The lighter green circles (r=89 mas) all have identical centers at WISPIT 2b (star-planet separation=309.43 mas, PA=242.2°). In the discovery image WISPIT 2b is detected in the ASDI image at a SNR=5.5 (top right), on April 16 it was recovered at SNR=12.5 (bottom right). The dashed yellow line in the Hα image traces the 2nd brightest ring (visible at z'; called ring #2; the brightest inner ring is ring #3; see Letter 1 for ring images and names). All images have the first 5 principal component (PC) modes removed by pyKLIP with movement set to zero. Images are 1518x1388 mas, smoothed by a FWHM=17 mas Gaussian (except the long-wavelength z' image, which has a 29 mas smoothing and pyKLIP movement=5, and a deeper stretch). North is up, and East is left in these, and all following, images.

*4.2. Follow-up of WISPIT 2b*

Once the location of the planet was known, it was recovered in archival (2023-10-19 and 2024-10-04) H band VLT/SPHERE IRDIS data (see Letter 1 for details). The very red protoplanet was recovered at higher SNR in deep Ks ESO-DDT images taken with VLT/SPHERE on 2025, April 26 (see Letter 1). We also obtained deep L' LBT-DDT images with the LBT with both 8.4m apertures (unstacked, dual beam) on the LMIRcam detector which is part of the LBTI instrument. The H and Ks images of WISPIT 2b are fully described in Letter 1. The LBT/LMIRcam L' observations are briefly described in the next section.

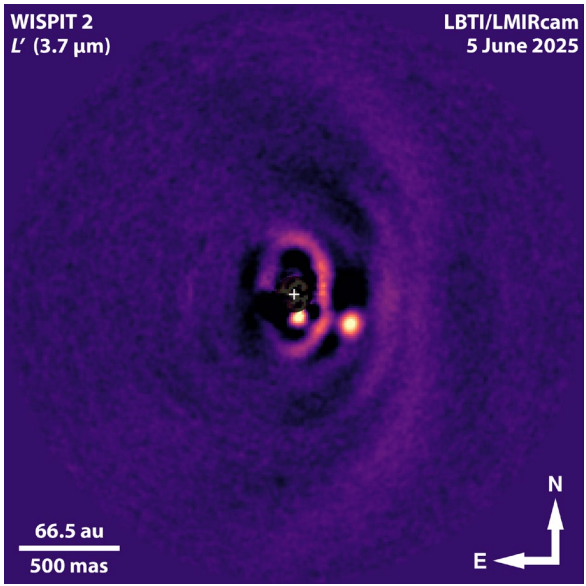

**Fig 2:** The KLIP reduced L' image from the LBT telescope with the LBTI LMIRcam instrument. The location of WISPIT 2b and the inner #3 and outer #2 dust rings are clear (despite the significant self-subtraction from KLIP). WISPIT 2b is located nearly in the center of the dust free gap between the rings. Image is 1.43x1.43" see Appendix B for details about the L' reduction.

*4.3. L' Follow-up of WISPIT 2b at LBT*

We observed WISPIT 2b at L' (3702 nm) at the LBT with the dual beam but unstacked mode of LBTI/LMIRcam (S. Ertel et al. 2020; J. Leisenring et al. 2012). These 2025, June 5 (UT) data were taken in sub-arcsec seeing to slightly over 1 arcsec seeing and photometric. The total amount of integration time was 3.15 hours. Only one (DX) of the 2x 8.4m telescopes reached very high (>90%) Strehl, throughout so we just reduced the DX side (single 8.4m scope). The L' band data was reduced in the usual manner for nodding L' ADI (G. Weible et al. 2025). The KLIP reduced image can be seen in Fig. 2. Please see appendix B for a complete description of the all observation and reduction steps for the L' data.

5.0 PHOTOMETRY AND ASTROMETRY

As PSF subtraction algorithms (like pyKLIP) can distort planet signal, we obtained companion astrometry and photometry through the forward modeling technique with the forward modeling feature in PyKLIP (J. Wang et al. 2015) for accurate measurements and uncertainties on the planet contrasts. For the astrometry we previously observed an astrometric calibration field in Baade's Window (J. Long et al. 2024) which has been extensively used by MagAO and GPI. We adopt for science camera 2 (the Hα camera, used for the astrometry in this work) a platescale of 0.00593±0.00003"/pix, PA$_{offset}$=+2.0±0.2 deg for this work, which is almost identical to that of Close et al. (2025) which, in turn, was in excellent agreement with the VLTI/gravity orbits of PDS 70 b and c (J. Wang et al. 2021, 2022).

| 2025 Hα data from Magellan/MagAO-X | 13 April; WISPIT 2b | 16 April; WISPIT 2b |
|---|---|---|
| Observed Hα Separation (mas) | **309.90±1.60** | **309.43±1.56** |
| Observed Hα PA (deg) | **242.71±0.82** | **242.21±0.41** |
| Forward modeled contrast results for 2b | | |
| ASDIcontrast$_{Hα}$: (flux of planet in ASDI image) / (Hα flux of star) | **(7.0±0.9) x 10$^{-4}$** (SNR=5.5) | **(6.5±0.5) x 10$^{-4}$** (SNR=12.5) |
| Hα line flux of planet 2b $f_{Hα}$ (erg/s/cm$^2$) | (1.38±0.33) x 10$^{-15}$ | (1.29±0.28) x 10$^{-15}$ |
| Planet 2b accretion rate $\dot{M}_p$ (Av=0-3 mag extinction) | 2.31$^{+3.85}_{-0.21}$ x 10$^{-12}$ M$_{sun}$/yr | 2.25$^{+3.75}_{-0.17}$ x 10$^{-12}$ M$_{sun}$/yr |

**Table 1:** The discovery Hα Photometry and Astrometry for April 13 and 16, 2025 (UT) for protoplanet WISPIT 2b. Values in **bold** text are directly measured, otherwise they are calculated values.

Our approach to photometry and astrometry of planet WISPIT 2b was straightforward. We used the fully forward modeled planet insertion option of pyKLIP to inject fake negative planets at the separations of b into each of the input images. Since we did not use a coronagraph for any of these observations, it is straightforward to measure the relative flux of the star and planet to obtain an accurate contrast. We fit a Gaussian to the final multi-hour "deep" WISPIT 2A PSF core (the PSF is the median of all selected images –all linear) to find accurate sub-pixel stellar peak counts and planet FWHM to model the fake planets as accurately as possible in pyKLIP. This, in turn, leads to the most accurate contrasts and astrometry from pyKLIP forward modeling.

| 2025, April 16: z' data from Magellan/MagAO-X | WISPIT 2b | CC1 |
|---|---|---|
| Observed z' Separation (mas) | **312±5** | **109.7±2.9** |
| Observed z' PA (deg) | **242±1** | **192±1** |
| Forward modeled contrast results from photometry | | |
| Contrast$_{z'}$: (z' flux of planet) / (z' flux of star) | **(9±5)x10$^{-6}$** (SNR~2) | **(2.4±0.5)x10$^{-4}$** (SNR=4.3) |
| Adopted magnitude of the star WISPIT 2A at z' | 10.46 mag (C. Onken et al. 2024) | |
| Magnitude of planet at z' | ~23.1 | 19.40$^{+0.65}_{-0.26}$ mag |
| Absolute Magnitude (Mz') of planet | ~17.4 | 13.78$^{+0.65}_{-0.26}$ mag |
| Est. Mass of planet from z' with DUSTY00 | 6±2 M$_{jup}$ at 5.1$^{+2.4}_{-1.3}$ Myr | 10±1 M$_{jup}$ at 5.1$^{+2.4}_{-1.3}$ Myr |

**Table 2:** The z' Photometry and Astrometry for 2025, April 16 (UT) for WISPIT 2b and CC1. Values in **bold** text are directly measured, otherwise they are calculated values.

We then followed L. Close et al. (2025) to run a grid search to add the perfectly centered and scaled fake negative planet to completely cancel (integrated flux in planet aperture = zero) the H$\alpha$ planet in the ASDI image. This yielded the ASDI$_{contrastH\alpha}$ = (7.0±0.9)x10$^{-4}$ on April 13 and ASDI$_{contrastH\alpha}$ = (6.5±0.5)x10$^{-4}$ on April 16, 2025 (see Table 1). Errors in the contrast were estimated from injecting a "ring" of 6 fake planets at the 309.5 mas separation and then using aperture photometry to estimate the standard deviation of planet photometry (Close et al. 2025).

| 2025, June 5: L' data from LBT LBTI/LMIRcam | WISPIT 2b | CC1 |
|---|---|---|
| Observed L' Separation (mas) | **315.9 ± 5.9** | **113 ±14** |
| Observed L' PA (deg) | **242.16±0.83** | **191.9 ± 2.4** |
| Forward modeled contrast results from photometry | | |
| Contrast$_{L'}$ : (L' flux of planet) / (L' flux of star) | **(1.76±0.07) x 10$^{-3}$** | **(2.80±0.37)x10$^{-3}$** |
| Magnitude of the star WISPIT 2A at L' | 8.42±0.02 mag (WISE) | |
| Magnitude of planet at L' | 15.30±0.05 mag | 14.80$^{+0.76}_{-0.43}$ mag |
| Absolute Magnitude (ML') of planet | 9.67±0.05 mag | 9.18$^{+0.76}_{-0.43}$ mag |
| Est. Mass of planet from L' with DUSTY00 | 5.3±1.0 M$_{jup}$ at 5.1$^{+2.4}_{-1.3}$ Myr | 8±4 M$_{jup}$ at 5.1$^{+2.4}_{-1.3}$ Myr |

**Table 3:** The L' Photometry and Astrometry for 2025, June 5 (UT) for WISPIT 2b and CC1. Values in **bold** text are directly measured, otherwise they are calculated values.

We did not significantly detect any point source at SNR≥5 in the excellent z' dataset at the location of WISPIT 2b (lower left, Fig 1). However, using a deeper stacking to 64x 120s images (pyKLIP movement=5) does reveal a z'~23 mag source at (9±5)x10$^{-6}$ contrast at WISPIT 2b's location, but only at a SNR~2. Which, if real, suggests a DUSTY model mass of 6±2 M$_{jup}$ (Table 2) from the z' flux. We caution that this is a very weak signal and so this z'~23 mag ($\Delta z'$=12.6 mag; 9x10$^{-6}$ contrast at 312 mas) source should not be considered a definitive detection without future confirmation. However, at H$\alpha$ the SNR>12.5 so the planet is real.

We note the faint extended z' "arc" in between the bright ring #2 and the fainter ring #3 is due to significant pyKLIP ADI self-subtraction of the edges of the bright rings, carving negative residuals. This well-known "ringing" ADI self-subtraction of the bright disk features leads to low-spatial noise artefacts in the otherwise dark gap.

A similar forward modeling procedure was followed for the L' datasets (see Appendix B). From forward modeling (as detailed in appendix B2) we find Contrast$_{L'}$ = (1.76±0.07) x 10$^{-3}$. We interpolate the WISE absolute photometry to derive a L' flux of 8.42±0.02 mag for WISPIT 2A, hence the magnitude of WISPIT 2b is 15.30±0.05 mag and so ML' = 9.67±0.05 mag. All the L' astrometry and photometry is reported in Table 3.

## 6.0 ANALYSIS

### 6.1. Example Hα Line Luminosity Calculation for WISPIT 2b April 16, 2025

The $L_{H\alpha}$ luminosity can be calculated for a gap planet of an extinction corrected effective "r' mag" at Hα (which we call r'mag_p_Hα) by comparing the its flux with Vega:

$$L_{H\alpha} = 4\pi D^2 f_{H\alpha} = 4\pi D^2 vega_{zeropoint\_r'} \Delta\lambda \{10^{\frac{r'mag\_p\_H\alpha}{-2.5}}\} \quad (1)$$

Where $f_{H\alpha}$ is the Hα line flux and r'mag_p_Hα is just the effective de-extincted "r' magnitude" w.r.t. Vega for planet "b" at Hα.

It is then necessary to tie the photometric system from the Hα flux of WISPIT 2A (which is too variable at Hα) to the continuum flux of WISPIT 2A; so we to need calculate: ASDIcontrast$_{continuum}$ = Flux_Hα/StarFlux_Cont . In other words, we need to compare to the 668nm continuum (in the r' band) since it is steady with time compared to Hα. The 4$^{th}$ data release of the Skymapper survey finds $r_A$=11.09 mag (which converts to $r'_A$=11.07 mag) C. Onken et al. 2024. We can estimate the stability of this r' flux by looking at the range of r' measurements over time ($r'_A$=10.80 mag AAVSO catalog (A. Henden et al. 2016) and $r'_A$=11.196 mag (C. Wolf et al. 2018); $r'_A$=11.20 mag Huang et al. 2022). Yielding a mean value of 11.06±0.22 mag consistent with the 11.09 mag from the final release of Skymapper (C. Onken et al. 2024). Hence, we adopt for WISPIT 2A: $r'_A$=11.1±0.2 mag.

To solve for ASDIcontrast$_{continuum}$ from our observations is given in Close et al. (2025) as:

ASDIcontrast$_{continuum}$= ASDIcontrast$_{H\alpha}$* β

Where, β = StarFlux_Hα / StarFlux_Cont * EMgain_CONT/EMgain_Hα * QE$_{CONT}$/QE$_{H\alpha}$

Where all of the parameters of β are easily measured ratios (all are listed in Table A1). The fact that β is completely dependent on ratios minimizes systematic errors which simply divide out in each ratio. Therefore, we can use the above relation to write equation 2:

ΔmagASDI$_{continuum}$ = -2.5*log10(ASDIcontrast$_{continuum}$) = -2.5*log10(ASDIcontrast$_{H\alpha}$ * β)  **(2)**

Since our 2025, April 16 ASDIcontrast$_{H\alpha}$ is (6.5±0.5) x $10^{-4}$ (Table 1), and β =0.29 (Table A1) therefore from equation 2 we know ΔmagASDI$_{continuum}$ = 9.31±0.12 mag. There is also a very slight correction since there is extra ~0.05 mag added due to Hα light in r' filter mag. So, the Hα flux projected as an "r" mag" of the planet is:

r'mag_p_Hα = (r'$_A$-A$_{r'}$)+(ΔmagASDI$_{continuum}$+0.05)-A$_p$ = (11.1±0.2-A$_{r'}$)+(9.31±0.12+0.05)-A$_p$ **(3)**

= 20.46±0.32 mag

So WISPIT 2b has an Hα line flux similar to an object with an r' flux 6.7x$10^{-9}$ fainter than Vega. In the case of the slowly accreting WISPIT 2b we have very little extinction to the star and planet (G. Marleau et al. 2022) and so we will assume A$_{r'}$=A$_p$=0 (K. Wagner et al. 2018; T. Thanathibodee et al. 2019; Y. Zhou et al. 2021) so equation 3 suggests an effective r'mag_p_Hα flux of b at Hα is similar to a continuum 668nm (Δλ=8nm) source with an r'~20.46 mag flux. Therefore, the line luminosity L$_{Hα}$ can be written:

$$L_{H\alpha} = 4\pi D^2 Vega_{zeropoint\_r'} \Delta\lambda \{10^{\frac{r'mag\_p\_H\alpha}{-2.5}}\} \quad (4)$$

Which we can directly solve for in the case of WISPIT 2b as:

Log(L$_{Hα}$/L$_{sun}$)=log(4π(133*3.1x$10^{18}$)$^2$*2.4x$10^{-5}$*0.008)/[3.9x$10^{33}$*$10^{((20.46±0.32)/2.5)}$]= -6.15

where the Vega zero point magnitude of the r' filter (Vega$_{zeropoint\_r'}$; J. Males 2013) is 2.4x$10^{-5}$ ergs/(s cm$^2$ μm). Since we are comparing the Hα flux to that of WISPIT 2A in the continuum filter we use Δλ=0.008 μm for our continuum filter. This Log(L$_{Hα}$/L$_{sun}$) = -6.15 is a significant amount of accretion emission at Hα, which allows for the very solid (SNR=12.5) detection of the protoplanet at Hα.

To calculate the Hα line flux (f$_{Hα}$) of b is simple, just divide the line luminosity L$_{Hα}$ by $4\pi D^2$ as is clear from equation 1. Therefore, the Hα line flux can be shown to be (1.29±0.28)x$10^{-15}$ erg/s/cm$^2$ on 2025, April 16 with a full and rigorous Gaussian error analysis of equation 4. See L. Close et al. (2025) their Appendix B and Fig. B2,for our Gaussian error propagation technique to produce a rigorous flux error estimate. In this manner, all the Hα line fluxes and errors in Table 1 were calculated.

In figure 3 we plot all the known Hα fluxes of all known Hα protoplanets with this uniform methodology. So that we can compare the different protoplanets we normalize all the fluxes at a distance of 113pc (that of PDS 70). From Fig. 3 we see the line flux of WISPIT 2b is very similar to all other Hα protoplanets.

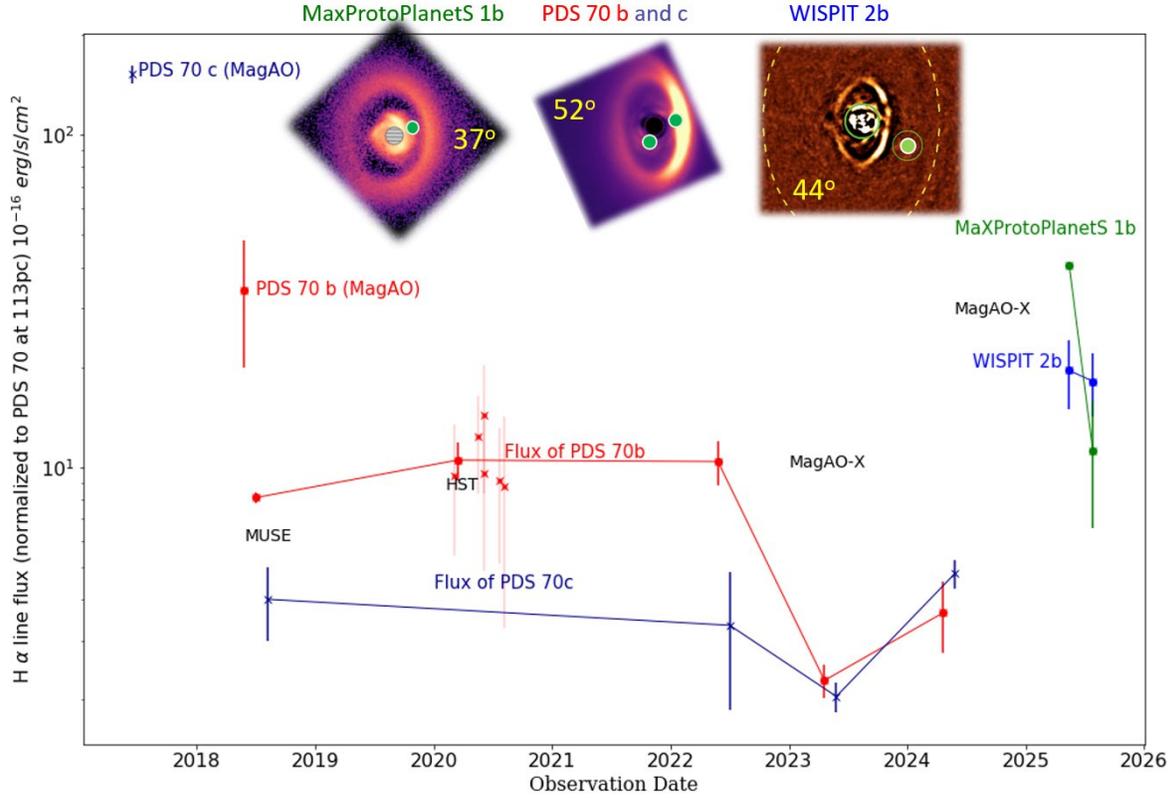

**Fig 3:** The calculated Hα line fluxes of all known Hα protoplanets, assuming no extinction ($A_R=A_P=0$ mag). They have been normalized to 113 pc (the distance to PDS 70). This covers 8 years of observations and covers all known Hα observations from literature. We include rotated and scaled thumbnail insets in scattered light of each of the systems with well-known Hα protoplanets. The green circles each represent the location of the Hα planet in that system. While the sample to date is very small, notice how all these systems have similar inclinations. Note how the full day side of these protoplanets is never seen. Inset disk images: MaXProtoPlanetS 1's disk in polarized scattered light from C. Ginski et al. (2025); PDS 70's disk in polarized light from Z. Wahhaj et al. (2024); and WISPIT 2's disk in z' from this work. Inclinations fit to the ALMA continuum disk images are called out in yellow text on insets.

*6.2. Mass Accretion Calculation for WISPIT 2b*

The planet mass accretion rates ($\dot{M}_p$) were calculated as in L. Close et al. (2025) in section 6.2 of that paper which used an semi-empirical (E. Rigliaco, et al. 2012) model of magnetospherical accretion (T. Thanathibodee, et al. 2019) to estimate the mass accretion rate required for the observed line fluxes (L. Close 2020). We adopt here a mass of 4.9 $M_{jup}$ consistent with the masses from Tables 2-3 based on the mass estimate of 4.9 $M_{jup}$ from H and Ks planet fluxes in Letter 1 (this value is further discussed in section 7.1 below). We adopt a radius of the planet of 1.6 $R_{jup}$ from the DUSTY evolutionary models. In this manner, the mass accretion rates in Table 1 were calculated.

We would like to compare WISPIT 2b's $\dot{M}_p$ values to all the other known protoplanets. We plot these in Fig. 4. Again we see that WISPIT 2b's $\dot{M}_p$ values are in-line with those of the other protoplanets. We note that in the published version of L. Close et al. (2025) there was a 24x scaling error so that $\dot{M}_p$ was too low by 24x for PDS 70 b and c. The $\dot{M}_p$ values in this letter (in Fig. 4) should be used instead for PDS 70 b and c.

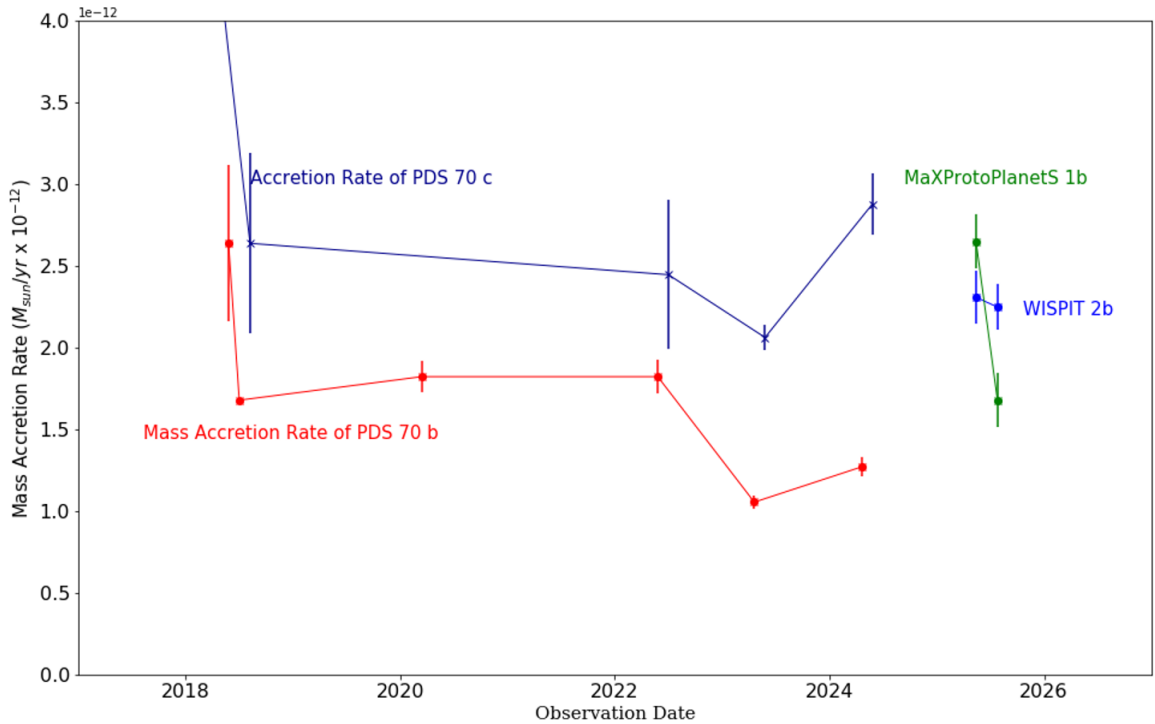

**Fig 4:** Similar to Fig. 3 except now we plot the estimated mass accretion rates onto these planets (assuming no extinction). As can be seen these are all currently accreting weakly (<4x10$^{-12}$ $M_{sun}$/yr or <0.004 $M_{jup}$/Myr). Which, in turn, suggests that the Hα self-absorption by the infalling gas should be weak, and so our assumption of there being no absorption/extinction of the Hα is reasonable according to the models of G. Marleau et al. (2022). We see that WISPIT 2b has a similar mass accretion rate to all the other protoplanets. For clarity some of the datapoints have been slightly shifted in time.

## 7.0 DISCUSSION
### 7.1. On the Age and Mass of WISPIT 2b

The star WISPIT 2A belongs to the young group Theia 53 which has an upper limit age of ~13.6 Myr (Letter 1). This upper limit makes good sense for WISPIT 2A since there is still a gas rich disk around WISPIT 2. We observed $Star_{H\alpha\_ph}/Star_{CONT\_ph}$ =0.43 (for reference, 0.125=no H$\alpha$) which implies strong EW(H$\alpha$, 6563)= -40.5 Å in emission measured by MagAO-X on 2025, April 13. This strong stellar H$\alpha$ emission implies WISPIT 2A is an actively accreting CTTS star since |EW(H$\alpha$, 6563)|>13 Å emission which is above the maximum expected stellar activity level for this star (Barrado y Navascues & Martin 2003).

We also obtained a LBT/PEPSI (K. Strassmeier et al. 2015) high-resolution R~250000 spectrum on 2025, June 19 (UT) and measured a lithium equivalent width of EW(Li, 6707)=0.35 Å, indicating the strong presence of Lithium and providing additional evidence that the star is indeed young and consistent with membership in Theia 53. A full analysis of the WISPIT 2A PEPSI spectra is beyond the scope of this letter and will be presented elsewhere. Our measured lithium and H$\alpha$ strength, in concert with other arguments (see Letter 1), allow us to adopt an age of $5.1^{+2.4}_{-1.3}$ Myr for WISPIT 2.

From our adopted $5.1^{+2.4}_{-1.3}$ Myr age range we find a mass range for the planet $5.3\pm1.0$ $M_{jup}$ from the DUSTY evolutionary models (I. Baraffe et al. 2002) applied to our L' photometry in Table 3. This is consistent with the WISPIT 2b mass of $4.9^{+0.9}_{-0.6}$ $M_{jup}$ found from H and Ks photometry in Letter 1. It is worth noting that the age of the planet could be younger than the star, and so these masses could be considered an upper limits. On the other hand, the age of any one star is very hard to definitely measure so ages/masses could be higher than our estimate, but as part of Theia 53 it must be less than ~13.6 Myr (Letter 1). So even in the extreme oldest case of 13.6 Myr yields an DUSTY model L' mass ~9.5 $M_{jup}$, hence the planet's mass is still within the planetary mass regime even at this large age. We may also be slightly (~10%) overestimating the mass of the planet from the L' flux alone. The slightly higher $M_p$=5.3±1.0 $M_{jup}$ mass derived at L' is 0.4 $M_{jup}$ higher compared to that estimated at H and Ks (Letter 1) might be due to a bit of excess L' light from a warm CPD dust component warmed by the planet itself. But the fact that the L', H and Ks model masses are all very consistent within the errors ($\Delta M_p$= $0.4^{+1.5}_{-1.3}$ $M_{jup}$) suggests that the CPD thermal L' emission is a relatively minor component of the L' light from WISPIT 2b, and so most of the L' light is likely from the planet's photosphere.

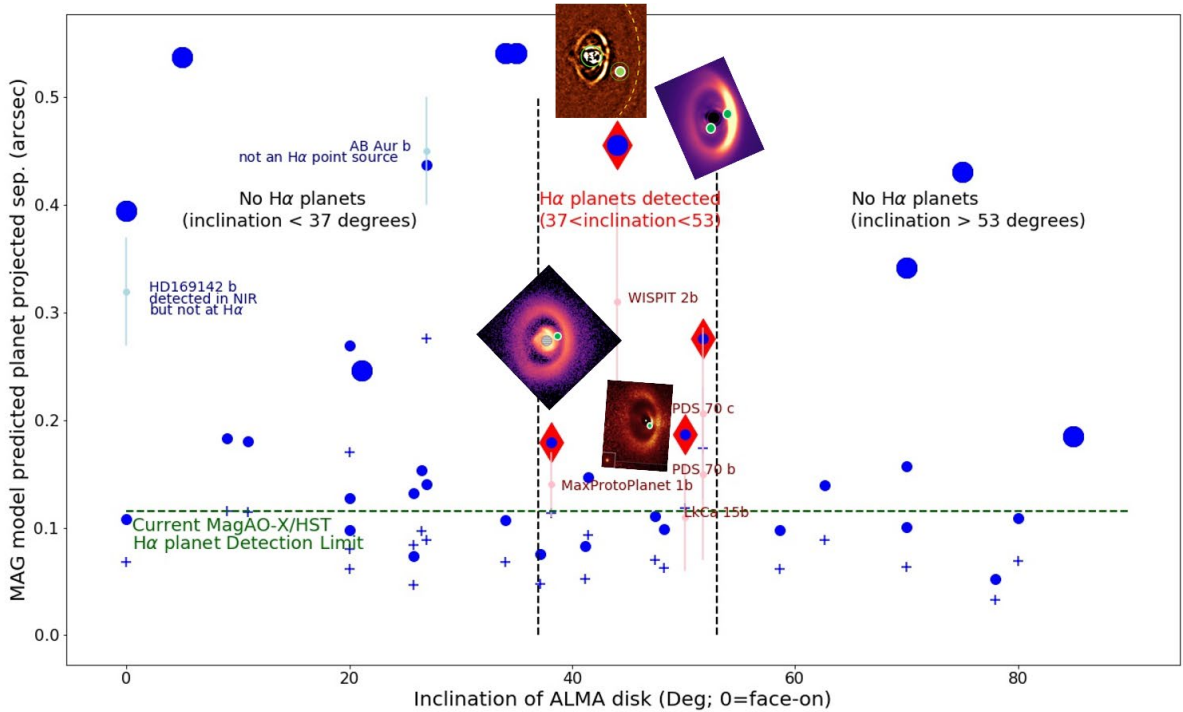

**Fig 5:** Here we plot predicted planet locations for all 33 known ALMA disks that have an inner cavity or a wide gap > 30 au. We use the MAG model (Close 2020) to predict the average time integrated projected separation of the massive planets that could be responsible for clearing the large gaps in all 33 of these systems. The small blue dots are the predicted locations of the outer ($a_2$) planet from the size of the cavity ($R_{cav}$) of each ALMA disk. The small blue "+" are the predicted locations of the inner planets ($a_1$) in 2:1 MMR with the outer planet in the cavity (L. Close 2020). The large blue dots are the predicted positions of the annular gap clearing planets. WISPIT 2b is the first and only known accreting H$\alpha$ annular gap planet, however, the plot also shows the inclinations of the many other ring gaps (large blue dots). Note how there are, to date, only H$\alpha$ planets (red diamonds) found in the zone: 37 < inclination < 55°. Despite the abundance of excellent targets (any blue symbol above the green dashed line at 115 mas is considered a "detectable" H$\alpha$ planet, but only those with red diamonds are actually detected), there are no H$\alpha$ planets detected with inclinations < 37° (pole-on systems). Nor are any found with inclinations > 55°. All known protoplanets are plotted as light dots with vertical lines (these vertical errorbars denote the estimated angle between the time averaged positions of the MAG model and a pole-on orbit). The red light dots are all the known H$\alpha$ protoplanets, the blue ones are known NIR planets that seem to have no direct H$\alpha$ emission.

*7.2. Are Some Inclinations Preferred for the Detection of H$\alpha$ Protoplanets?*

While for many years the only *bona-fide* H$\alpha$ Protoplanets were PDS 70 b and c, we now have the addition of MaXProtoPlanet 1b (J. Li et al. 2025) and WISPIT 2b, so now is a good time to look at this small, but important, exoplanet population. The first H$\alpha$ protoplanet detection was LkCa 15b (S. Sallum et al. 2015), where later the sparse aperture masking continuum "detections" of LkCa 15 b, c and d were shown to be, in fact, tracking dust clumps in the inner disk (T. Currie et al. 2019). However, the SNR=6 Hα detection of LkCa 15b still seems robust, and there have been recently other (L. Close et al. 2025b) Hα detections, so in this section we will also consider LkCa 15b as an Hα protoplanet candidate.

It is interesting to note that the inclinations of these ALMA disks in continuum are 37º, 44.0º, 50.1º, 51.7º (MaXProtoPlanetS 1b (J. Li et al. 2025); WISPIT 2b; LkCa 15b; PDS 70 b and c; respectively –see Fig 3). This group is heavily clumped 37≤$i$≤52º and is *a priori* unlikely to be selected from a random drawing of 4 inclinations from 0 to 90º. However, the sample size is small, and this could be just a strange coincidence, or a consequence of selection effects in the parent population of transitional disks.

In figure 5 we try to understand this parent population of ALMA large (>30 au) cavity or annular gap (> 30 au) disks that have been investigated at high-contrast Hα by AO (G. Cugno et al. 2019; A. Zurlo et al. 2020; N. Huélamo et al. 2022) and/or *HST* (Y. Zhou et al. 2022; 2025). We see from Fig. 5 that for inclinations < 37º there are 15 "detectable" planets. We define a detectable planet as a gap clearing Hα planet predicted by the MAG model (L. Close 2020) to have separations>115 mas (horizontal line in Fig. 5) based on the observed ALMA continuum gap size. Of these 15 hypothetical planets (carving the gaps in these $i$<37º disks) none are detected at Hα, so detection rate = 0/15 =0%.

Two of these low inclination ($i$<37º) stars had candidate planets that were detected in the NIR (but not at Hα); AB Aur b and HD169142 b. Planet candidate HD169142 b (I. Hammond et al. 2023) is right in the cleared "ring" gap (dark annular gap between 2 dust rings), but despite deep Hα searches from MagAO, MagAO-X, and SPHERE/ZIMPOL there has never been any Hα excess confirmed from this nearly pole-on ($i$~0º) system and none from HD169142 b itself (B. Biller et al. 2014; and Letter 1), so we consider it a "non-Hα planet" candidate. The other low inclination protoplanet candidate is AB Aur b, but it is not even point-like, instead it is an extended dusty "clump", that might be a protoplanet still embedded in its CPD, with no direct view of the Hα line formation region found after an extensive *HST* Hα imaging campaign (Bowler et al. 2025), hence also a "non-Hα planet".

For with $i$ >52º there are 5 potentially detectable planets in Fig. 5 (where the bias against high inclination systems is properly considered; Close 2020). We also see from Fig. 5 that none of these higher inclination systems have Hα planets. Hence, 0/5=0% detection rate for $i$ >52º.

On the other hand, as is clear from Fig. 5, in the range 37≤*i*≤52º we find a possible 5 systems with predicted planets>115 mas. Of those 5 we find Hα protoplanets around 4 of them (red diamonds in Fig. 5), so a detection rate of 4/5=80%.

Only one star (HD 97048) is within this range and doesn't have an Hα planet. HD 97048 is at *i*=41.4º, but despite a ring system (C. Ginski et al. 2016; 2024) similar to WISPIT 2, it doesn't have a directly imaged Hα protoplanet (N. Huélamo et al. 2022 and reference within) nor any directly imaged planets. However, it has been reported that there is a planet (a few Jupiters in mass) in the outer gap, detected as an ALMA CO "velocity kink" protoplanet (C. Pinte et al. 2019), but this planet has not been directly detected at any wavelength yet.

This 80% success rate is a high rate for the direct detection of any class of exoplanets, such searches usually have planet direct detection rates of just few percent or less. For example, the massive Gemini/GPIES survey targeted 531 stars and had only a few detections. GPIES was designed to select "blindly" from a carefully selected survey of the nearby young (≤0.5 Gyr) star population where GPIES discovered that only $9^{+5}_{-4}$% of stars have 5-13 $M_{jup}$ exoplanets at 10-100 au (E. Nielsen et al. 2019). In contrast, we are targeting the largest (>30 au) gap transitional disks forcing these gap planets to be >115 mas separation (horizonal line in Fig. 5). So it is not that surprising that we find massive (2-8 $M_{jup}$), wide (20-60 au) Hα planets in the cavities and gaps of these large disks. But it is somewhat surprising that they are all (so far) in the $37 \leq i \leq 52$º range and that 80% of our targets in this range have detected planets –and that none of the planet candidates outside this range have any Hα detected.

Our success rate does not, in any way, invalidate the exoplanet population distribution results of E. Nielsen et al. (2019) –wide massive exoplanets are still rare. We are just greatly improving our chances of exoplanet discovery by directly targeting wide gap transition disk stars that *a priori* should have "detectable" massive wide Hα planets clearing these gaps (S. Dodson-Robinson & C. Salyk 2011; L. Close 2020). Our work does demonstrate that Hα ASDI can be a very powerful observational technique for discovering new protoplanets in wide gaps (at least in the 37≤*i*≤52º range). Indeed, all the well-known Hα protoplanets (PDS 70c, MaxProtoPlanetS 1b, WISPIT 2b) were all first discovered at Hα (save PDS 70 b, which was first detected at H and Ks; M. Keppler et al. 2018).

We can ask what are the *a priori* odds that all 4 Hα planets share such a clustered inclination distribution (37≤*i*≤52º) selected randomly from our "detectable" 25 large gap disks in Fig. 5. In a series of $10^6$ simulations of the dataset of Fig. 5, where we generated a similar group of 25 detectable planets (all "detectable" with separation>115 mas –separations based on the (as yet unproven) presence of gap clearing planets; S. Dodson-Robinson & C. Salyk 2011; L. Close 2020) and selected 4 randomly to be Hα planets, only 3.315% of the time did all 4 Hα planets cluster around any possible $\Delta i \leq 15°$ ranged cluster. The probability drops further to 1.076% (2.6σ) if we also impose the observed ≥80% detection rate (which means only 0 or 1 other of the 21 remaining "non-Hα planets" allowed inside the $\Delta i$ range) over this cluster (as we observed in Fig. 5). So, observationally speaking, inclination may appear to be important to the success at detection of Hα emission. It appears that inclinations in the range $37 \leq i \leq 52°$ are preferred. But we caution that this is still a small sample, just 4 Hα systems out of a possible 25 "detectable" targets, and that more Hα systems need to be found and/or searched in the future to increase (or decrease) the 2.6σ significance of this finding. Our 2.6σ significance is less than 3σ and so not yet statistically significant with this small sample size of 4. In appendix C, we speculate about one possible toy model of why the Hα produced by magnetospherical accretion onto a protoplanet might naturally be sensitive to the inclination of the line of sight.

### 7.3. The Search for Other Planets in the WISPIT System

Given the excellent datasets that we obtained it is logical to carry specialized data reductions aimed at the detection of faint outer and inner planets. We carried out an extensive suite of different reductions with different pyKLIP parameters, clock induced charge (CIC) mitigation, and cosmic ray rejection algorithms, and a wide range of high pass filters. In the end, there were no other significant (SNR>2) Hα emission point sources found in any of our datasets. Only WISPIT 2A (the star) and planet 2b had significant Hα emission on 2025, April 13, 16.

*7.3.1 A Close Companion: CC1*

There is an intriguing inner point-like object inside the inner cavity. In Fig. 6, we show that there is a Close Companion candidate "CC1" (Separation=110 mas (15 au deprojected); PA=192°) first detected at z' (908 nm) (SNR~4) with MagAO-X and then at L' (3702 nm) with LMIRcam (SNR>12). For CC1 we measure forward modeled point-source contrasts of $(2.4\pm0.5)\times10^{-4}$ at z' (z'=$19.40^{+0.65}_{-0.26}$ mag; Table 2) and $(2.80\pm0.35)\times10^{-3}$ at L' ($14.80^{+0.76}_{-0.43}$ mag; Table 3). In contrast to z' and L', we do not have high SNR detections of CC1 at our other NIR bands (H and Ks) likely due to the large (d=185 mas) size of the (N_ALC+YJH_S) coronagraph used with SPHERE (Letter 1). Although there are hints of CC1 in total intensity (it appears unpolarized) in the 2025, April 26 "BB_Ks" SPHERE dataset of Letter 1 with "ADI" and "cADI" reductions (reproduced here; lower left of Fig. 6). We do not detect CC1 in the 668 nm continuum filter, nor do we have a significant Hα detection (see Fig. 6 upper left).

*7.3.2 On the Nature of CC1: A Dust Clump or an Inner Planet?*

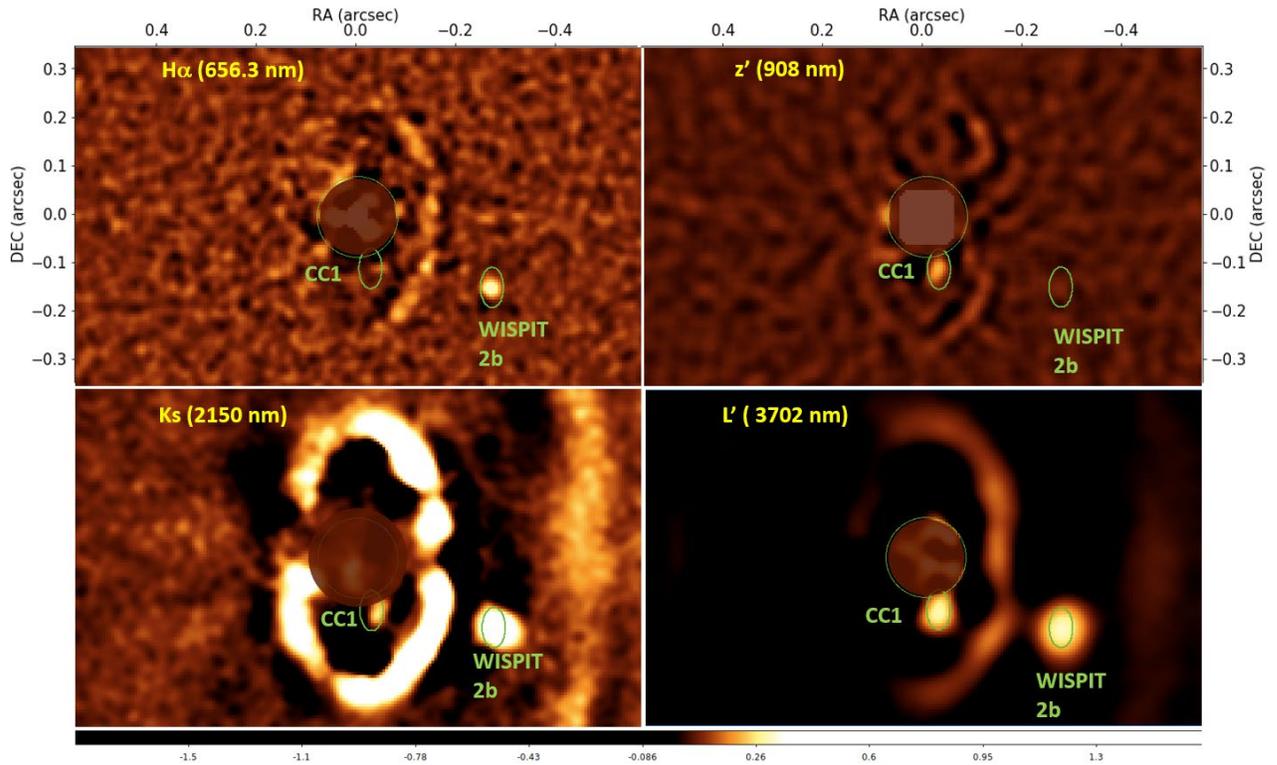

**Fig 6**: Here we zoom in on the inner 1162x694 mas to examine the position of the Close Companion (CC1) at different wavelengths. The z' image is same as Fig 1, but shown here is with movement=0. The cADI Ks image is from SPHERE/IRDIS; see letter 1 for details. The L' image is the same as in Fig 2, but increased smoothing. This very red (z'-L'~4.6 mag) CC1 source has a SNR=4.3 in the z' image and SNR=12.1 in the L' image. There is little difference between the z' – L' positions (Δsep=-3±14 mas and ΔPA=0.1±3.0°) so they are consistent with co-spatial. The Ks source is partially blocked by the diffracted edge of the SPHERE r=93mas coronagraph (dark circle) yet is spatially consistent with the z' and L' sources within errors. CC1 is a point source at L' but appears slightly extended at z' and Ks. CC1 is not significantly detected in the Hα image (SNR<2). The CC1 green ellipse is in the same location (110 mas; 192°) in each image. The ellipses trace r=41 mas (5 au) circles with $i$=44° (same as main disk). There is a r=83 mas semi-transparent circle centered exactly on the star. It is unclear if CC1 is a warm inner protoplanet (mass ~9 $M_{jup}$) or an unusually red compact dust clump. If it is a planet, it could appear modesty extended due to its dusty CPD's edge scattering starlight (no z' scattering occurs at 2b due to the shadow cast by ring 3). Scaling from the observed size (~1 Hill sphere) of the PDS 70 b and c CPDs of Close et al. (2025) we predict CC1's CPD would appear elongated by r~2.7 au. The observed size of CC1 at z' is the PSF convolved with the CPD, yielding a CPD radius of ~24 mas at z'. Hence, CC1's CPD's edge should trace ~60% inside of the green CC1 ellipse, roughly consistent with the elongation observed in z'.

CC1 does appear slightly extended in Fig. 6 at z' (but a point source at L'), so it might be a dust clump. Moreover, Fig. 6 only shows a very weak Hα with a low SNR~2 and so is likely not significant. So it is not (at least on 2025 April 13,16) an Hα protoplanet –so it could be dust. On the other hand, it is very bright and point-like at L' and we find CC1 has a very red color (z'-L'=4.6 mag) which argues that CC1 could have a "warm" blackbody $T_{eff}$ ~1-2 x$10^3$ K. The dust at 15 au is much cooler than this, and so CC1 cannot solely be produced by a passively cooling dust clump at 15 au. Nor can its very red colors (z'-L'=4.6 mag) be produced by starlight gray scattering off compact dust. We might be able to explain CC1 as a very special dust clump that only reflects long wavelengths (shifting the starlight to a reddened z'-L'~4.6 mag). However, this would require very unlikely dichroic scattering dust grain properties. Indeed, the reddest known dust disk is that of HR 4796 which has an observed reflectance z'-L'~1.0 mag/arcsec$^2$ (H. Zhong et al. 2024) which gives a max red dust color of z'-L'~2.9 (considering the spectrum of WISPIT 2). Moreover, even the largest, reddest grains give only theoretical z'-L' colors of <3.0 mag (R. Tazaki et al. 2019). Hence, our z'-L'=4.6 mag color appears too red for CC1 to be solely explained by scattered light from dust. Moreover, CC1 was not detected in polarized light at Ks only detected in total intensity, also inconsistent with scattered light off dust.

Another explanation is that CC1 might possibly be a low mass star with a dusty edge-on disk inclined directly towards our line of sight. This could produce very red colors by extinction. However, the observed smooth "non-spiral" WISPIT 2 ring system would be impossible with such a massive inner stellar mass companion at 15 au. This is just too much tidal mass for CC1 to have so close to ring #3 and yet leave this ring completely unperturbed and symmetric.

If we look, instead, at the exoplanet hypothesis for CC1, we see the L' ($14.80^{+0.76}_{-0.43}$ mag) and z' ($19.40^{+0.65}_{-0.26}$ mag) flux of CC1 could be due to a warm dusty planetary photosphere (as is the case of WISPIT 2b). Such an hypothesis would predict that CC1 would be a 8±4 $M_{jup}$ planet from its L' flux and a 10±1 $M_{jup}$ planet from z' flux (Tables 2, 3; DUSTY00; I. Baraffe et al. 2002). In any case: 1) the z' and L' photometry; 2) non-detection at 668nm; 3) non detection in polarized light at H+Ks are all consistent in color and magnitude with CC1 being from a photosphere of a 9±4 $M_{jup}$ exoplanet (DUSTY00 model). The lack of significant Hα could be due to a drop in accretion (which is known to happen to Hα protoplanets; L. Close et al. 2025) and/or optical extinction from its spatially extended CPD.

At a deprojected separation of 15±1 au CC1 is consistent with the 14.4 au 8:1 MMR orbital resonance with WISPIT 2b at 57.5 au (Letter 1's orbit fit for 2b). Such an MMR could help create multi-planet orbital stability in the WISPIT 2 system (L. Close 2020). Moreover, the presence of a massive inner 9 Mjup planet could scatter out an outer planet, and so be the cause of having such a wide ~58 au outer 5 Mjup planet like WISPIT 2b (Smullen et al. 2016).

Future follow-up spectra at K with VLTI/gravity, or 3-4μm with LBTI/ALES (A. Skemer et al. 2015), tracking orbital motion and higher-contrast coronagraphic z' & H-band imaging could all help determine if CC1 is truly another protoplanet (a "WISPIT 2c"), or an unusually red (z'-L'~4.6 mag) compact dust clump, or perhaps a protoplanet still somewhat embedded inside its CPD.

## 8.0 CONCLUSIONS

We present nearly diffraction-limited (<25 mas) Hα images of the star TYC 5709-354-1 which was recently discovered by the WISPIT survey to have a large multi-ring transitional disk. For more details about WISPIT 2 see van Capelleveen et al. (2025) which is a companion paper (Letter 1) to this work. Our Hα images of 2025, April 13 and April 16 discovered an accreting protoplanet. This protoplanet WISPIT 2b was at r=309.43±1.56 mas, (~57.5 au deprojected assuming i~44° inclination co-planar orbit) PA=242.21±0.41 with a SNR=5.5 and 12.5 respectively. WISPIT 2b appears to be clearing a dust-free gap between the two bright dust rings in the system.

This is the first time an ring or annular gap Hα protoplanet (a protoplanet with Hα emission found between two bright and narrow dust rings; like rings #3 and #2) has been discovered. Previous Hα protoplanets (like PDS 70 b and c) have been in the large inner cavity in their star's transitional disk. Here WISPIT 2b appears to be clearing a dust-free gap between the two bright rings of dust –as long predicted by theory.

WISPIT 2b is an actively accreting Hα protoplanet. We find Hα ASDI contrasts of (7.0±0.9) x $10^{-4}$ and (6.5±0.5) x $10^{-4}$ and so calculate an Hα line flux of (1.38±0.33) x $10^{-15}$ and (1.29±0.28) x $10^{-15}$ erg/s/cm² on April 13 and 16 respectively.

We also present L' photometry from LBT/LBTI of the planet (L'=15.30±0.05 mag) with the LMIRcam camera which, when coupled with an adopted age of $5.1^{+2.4}_{-1.3}$ Myr, yields a planet mass estimate of $M_p$=5.3±1.0 $M_{jup}$ and size $R_p$=1.6±0.2 $R_{jup}$ from the DUSTY evolutionary models (I. Baraffe et al. 2002). Given the main parameters of the planet we utilized the methodology of L.

Close et al. (2025) and calculated that WISPIT 2b is accreting at $2.25_{-0.17}^{+3.75} \times 10^{-12}$ $M_{sun}$/yr. WISPIT 2b joins 3 other known H$\alpha$ accreting protoplanets. WISPIT 2b is very similar to these other H$\alpha$ protoplanets in terms of inferred mass (2-8 $M_{jup}$), age (~5-10 Myr), and H$\alpha$ line luminosity and estimated mass accretion rates (1-3 x$10^{-12}$ $M_{sun}$/yr).

We note that the inclination of the system (44º) is very similar to all the other known H$\alpha$ protoplanet systems: $i$=37º (MaXProtoPlanetS 1b); 50º (LkCa 15b); and 52º (PDS 70 b and c). We argue, despite the small sample size, that a clump of inclinations ($\Delta i \leq 15$º) (centered on any central value with an >80% H$\alpha$ planet detection rate) has only a 1.0% (2.6$\sigma$) probability of occurring randomly, and so we speculate that magnetospherical accretion might have preferred inclinations (37-52º) for the direct (cloud/dust free) line of sight to the H$\alpha$ line formation/shock regions on the planetary surface. Future studies are required to see if this inclination trend is really significant. Detailed future theoretical modeling is required to see if the line of sight to H$\alpha$ emission could/should be inclination dependent.

We detected a compact point-like object CC1 in L' at 110 mas (~15 au deprojected) at PA=192º. We also detected CC1 at z' and Ks at this same location. CC1 has very red colors (z'-L'~4.6 mag). It had no significant H$\alpha$ emission on 2025 April 13, 16. However, our z' and L' photometry is consistent in magnitude and color with a 9±4 $M_{jup}$ exoplanet (a WISPIT 2c ). Or it could also be an unusually red dust clump inside the central cavity of WISPIT 2. Future observations of CC1 will be required to ascertain its true nature.

ACKNOWLEDGMENTS

We would like to thank the anonymous referee whose careful reading of the manuscript and excellent suggestions led to a much improved final manuscript. Laird Close was partially supported by past NASA eXoplanet Research Program (XRP) grants 80NSSC18K0441 and XRP 80NSSC21K0397 which funded the MaxProtoPlanetS survey. Jialin Li is also supported by U.S. NSF Graduate Research Fellowship. We are very grateful for support from the U.S. NSF MRI Award #1625441 (for MagAO-X development). The MagAO-X Phase II upgrade program is made possible by the generous support of the Heising-Simons Foundation. The results reported herein benefited from collaborations and/or information exchange within NASA's Nexus for Exoplanet System Science (NExSS) research coordination network sponsored by NASA's Science Mission Directorate. This material is based upon work supported by the National Aeronautics and Space Administration under Agreement No. 80NSSC21K0593 for the program "Alien Earths".

*Facilities*: Magellan:Clay (MagAO-X); LBT (LBTI); LBT (PEPSI)

*Software:* Python (https://www.python.org/), SciPy (Virtanen et al. 2020), NumPy (Oliphant 2006), Matplotlib, astropy (Astropy Collb. 2013, 2018). The development of pyKLIP is led by Jason Wang and collaborators see https://pyklip.readthedocs.io/en/latest/. Vortex Image Processing (G. Gonzalez et al. 2017; Christiaens et al. 2023).

Corresponding author lclose@arizona.edu

[1] Center for Astronomical Adaptive Optics, Department of Astronomy, University of Arizona, 933 N. Cherry Ave. Tucson, AZ 85718, USA

[2] Leiden Observatory, Leiden University, PO Box 9513, 2300 RA Leiden, The Netherlands

[3] Leibniz Institute for Astrophysics Potsdam (LIAP), An d. Sternwarte 16, 14482 Potsdam, Germany

[4] Center for Computational Astrophysics, Flatiron Institute, 162 5th Avenue, New York, New York, USA

[5] LBT org, Steward Observatory, University of Arizona, USA

[6] School of Natural Sciences, Center for Astronomy, University of Galway, Galway, H91 CF50, Ireland

[7] The Earth and Planets Laboratory (EPL), Carnegie Institution for Science, USA

[8] Amherst College Department of Physics and Astronomy, Science Center, 25 East Drive, Amherst, MA, USA

[9] Wyant College of Optical Sciences, The University of Arizona, 1630 E University Boulevard, Tucson, Arizona, USA

[10] University of Montreal, Montreal Quebec, Canada

[11] Subaru Telescope, National Observatory of Japan, NINS, 650 N. A'ohoku Place, Hilo, Hawai'I, USA

[12] Astrobiology Center, National Institutes of Natural Sciences, 2-21-1 Osawa, Mitaka, Tokyo, Japan

[13] Draper Laboratory, 555 Technology Square, Cambridge, Massachusetts, USA

[14] Department of Astronomy, University of Michigan, 323 West Hall, 1085 S University Ave, Ann Arbor, MI 48109

[15] Starfire Optical Range, Kirtland Air Force Base, Albuquerque, New Mexico, USA

[16] Northrop Grumman in Rolling Meadows, Illinois, USA

[17] Department of Physics, National Taiwan Normal University, Taipei 116, Taiwan

[18] College of Arts and Sciences, Queen's University, Canada

**APPENDIX A:** The Hα MagAO-X observations

Table A1 is log of all the observations and settings for these WISPIT Hα observations.

| | 13 April 2025 | 16 April 2025 |
|---|---|---|
| *Environmental* | | |
| Seeing (") | 0.68-1.08" | 0.34-0.52" |
| Wind (mph) | 22.7-31.4; NNE | 4.9-14.6; "N" |
| Photometric sky? | yes | yes |
| *Adaptive Optics Settings of MagAO-X* | | |
| Number of AO Modes Corrected | Auto gain 1000 | Auto gain 1000 |
| AO Loop Speed (Hz) | 1000 | 1000 |
| NCP DM | BMC 1024 (1K) | BMC 1024 (1K) |
| NCP aberration Correction | FDPR | FDPR |
| *EMCCD Science Camera Features* | | |
| Camera 1 filter : $\lambda_1, \Delta\lambda_1$ (CONT) | 668.0, 8.0 nm | z' (908, 131nm) |
| Camera 2 filter : $\lambda_2, \Delta\lambda_2$ (Hα) | 656.3, 1.045 nm | 656.3, 1.045 nm |
| Bump Mask in pupil? | No, open | No, open |
| $EM_1$ (CONT) as set on camera 1 | 600 | 500 |
| $EM_2$ (Hα) as set on camera 2 | 900 | 900 |
| EMgain $_{CONT}$ (ADU/e-) | 137.52±0.97 | 114.60±0.77 |
| EMgain $_{Hα}$ (ADU/e-) | 294.13±0.29 | 294.13±0.29 |
| Readnoise$_1$ rms e- (CONT) | 0.16 | 0.19 |
| Readnoise$_2$ rms e- (Hα) | 0.05 | 0.05 |
| *Exposure Times and WISPIT 2 Hα Observational Parameters* | | |
| Exposure time (DIT) | 0.5s | $DIT_1$=0.25  $DIT_2$=1s |
| Percentage of raw frames kept | 64.35% | 99.5% |
| Number of Hα raw frames kept | 8941 | 8084 |
| Exposure time of combined images | 120x 0.5=60s | 60x 1=60s |
| #of combined images fed to pyKLIP | 74 | 130 |
| Total deep exposure time (hours) | 1.23 hr. | 2.16 hr. |
| ADI sky rotation (start→stop: Δdeg) | -130→-175: 45° | -127→+177: 56° |
| High-Pass (HP) filter value (pix) | 4.1 | 4.1 |
| StarFlux$_{Hα}$/StarFlux$_{CONT}$ | 0.624 | --- |
| QE$_{CONT}$/QE$_{Hα}$ | 16.8/16.6=1.01 | NA -SciBS 50/50 |
| r' mag of WISPIT 2A | 11.1±0.2 | 11.1±0.2 |
| FWHM of Hα PSF (deep image) | 25 mas | 23.6 mas |
| Strehl of Hα PSF (deep image) | 8-12% | 30% |
| ***Beta(β)*** = (StarFlux$_{Hα}$ / StarFlux$_{Cont}$)*(EMgain$_{CONT}$/EMgain$_{Hα}$)*( QE$_{CONT}$/QE$_{Hα}$) | | |
| Beta (β) | 0.29 | 0.29 |
| SDI "Contrast boost" = 1/β | 3.45x | 3.45x |
| ASDIcontrast$_{continuum}$ = ASDIcontrastHα * β = (ASDI planet flux)/(Star continuum flux) | b=(2.0±0.2)x10$^{-4}$ | b=(1.8±0.1)x10$^{-4}$ |
| *pyKLIP parameters and SNR* | | |
| pyKLIP Sectors, Annuli, PC Modes | 4, 10, 5 | 4, 10, 5 |
| pyKLIP movement | 0 | 0 |
| SNR of b in ASDI image | 5.5 | 12.5 |

**Table A1:** Log of all the observations, settings, and parameters used for our WISPIT 2b Hα observations.

**APPENDIX B:** LBTI/LMIRcam L' data reduction

*B1: L' Observations and Reductions*

TYC 5709-534-1 was observed with the Large Binocular Telescope Interferometer (LBTI: Hinz et al. 2016; Ertel et al. 2020) and the L/M-band InfraRed Camera (LMIRcam: Skrutskie et al. 2010; Leisenring et al. 2012) with an L′ filter ($\lambda$ cen = 3.702 $\mu$m, $\Delta \lambda$ FWHM = 0.584 $\mu$m) on 5 June 2025. The observations used the (pupil-stabilized) double-sided direct imaging mode with wavefront correction provided by the adaptive secondary mirrors of the LBT AO "SOUL" system (Esposito et al. 2010; Bailey et al. 2014; Pinna et al. 2016, 2023). The telescope was nodded in an ABAB pattern for modeling and subtraction of the background. Five non-destructive reads were collected "up the ramp" for each 2540.97 ms exposure, alternating nod positions every 250 exposures out of 5000 for a total exposure time of ~3.53 hr. > 62° of field rotation were observed, enabling ADI-based PSF subtraction (Marois et al. 2006).

A custom IDL+Python pipeline adapted from that described in Weible et al. (2025) was used for preprocessing; the following corrections are applied to each exposure. We performed a linear fit to the ramps after excluding saturated values. Poor fits ($r^2 < 0.8$) and negative slopes were masked. Standard calibrations (dark-subtraction, flat-fielding) were applied, and a bad-pixel mask was generated from the calibration frames. Bad pixels were interpolated with the nearest 24 good values. Vertical and horizontal striping was removed by subtracting the median of each column and row from itself after masking the stellar PSFs, and we iteratively sigma-clipped the images with 3−7-pixel 3.5-$\sigma$ filters. A PCA based subtraction of the thermal background was employed following Hunziker et al. (2018) and Rousseau et al. (2024), with the 1000 nearest frames taken at the alternate position used as references. We retained the top 10 PCs from the eigen decompositions and masked the target frames' PSFs before projection. These projection coefficients were used to reconstruct the complete background models from the unmasked eigenimages. The frames were de-striped a second time in the same manner as before and corrected for optical distortion from a warping solution provided by pinhole-grid observations (Maire et al.

2015; Spalding & Stone 2019). 331 × 331 sub-arrays centered on the PSF locations were cropped for post-processing.

The AO correction for the DX (right) pupil was found to attain a significantly higher Strehl ratio than that of the SX (left). The following processing is applied only to the DX images to prioritize recovering signals close to the diffraction limit of $\lambda/D \approx 90$ mas. The stellar PSFs were centered to sub-pixel precision with 2D Gaussian fits using the *cube_recenter_2dfit* Vortex Image Processing (VIP: Gomez Gonzalez et al. 2017; Christiaens et al. 2023) Python function. Centered images were separated into two semi-independent datasets for separate processing: one for each nod position. The inner regions of the frames of each dataset were correlated with their temporal median and the ~10% of images having the lowest Pearson correlation coefficients were trimmed with the *cube_detect_badfr_correlation* VIP function. The 4457 retained images (effective exposure ~3.15 hr) were centered a second time with 2D Gaussian fits. Residual background offsets measured in an annulus were subtracted, and each stellar PSF was normalized to its mean flux.

*B2: KLIP and Forward Modeling L' Photometry and Astrometry*

For our primary (forward-modeling) reduction the images were binned temporally by 1 (3) with a mean. We use the pca_annular VIP implementation of the KLIP algorithm with a parallactic angle (PA) threshold to model and subtract the stellar PSFs (Soummer et al. 2012; Amara & Quanz 2012; Absil et al. 2013). We reduce only the inner ~1.49" (~640 mas) of each frame for the primary (forward-modeling) reduction. The PA constraint for PSF references corresponds to an arclength $\approx$ 1 FWHM for planet b. We retain the top 9 PCs for subtraction and combine the residual images with an inverse-variance weighted trimmed mean (Bottom et al. 2017; Brandt et al. 2013). The reduced images from the two nodding positions were averaged. The result from our primary reduction is shown north-up in Figures 2 and 6 with a plate scale of $10.624^{+0.035}_{-0.038}$ mas/pixel as determined from routine astrometric observations of the Orion Trapezium.

Relative astrometry and L′ photometry of planet b and CC1 are found from iterative injections of scaled Gaussian PSF models into the two semi-independent datasets before PSF subtraction and combination. We used the *firstguess* VIP function to minimize the standard deviation of residuals in a FWHM-diameter aperture at the location of the point source. For ease of computation, the forward-modeled residual images are median-combined. The forward-modeling results are presented in Table 3. Uncertainties were computed as detailed in Weible et al. (2025), except that the standard deviations of aperture photometry and 2D Gaussian centroids for a given source (b, CC1) and two corresponding artificial injections at rotated PAs, i.e., of 6 measurements per source including both nod positions, were used to estimate measurement uncertainties.

**APPENDIX C:** Why Inclination Might Matter for the Observability of H$\alpha$

The heating effect of magnetospherical accretion on the upper cloud layers of protoplanets is an interesting topic that has not been extensively studied. While Z. Zhu et al. (2016); Y. Aoyama et al. (2018); T. Thanathibodee et al. (2019); Y. Aoyama et al. (2021); G. Marleau et al. (2022) all extensively study the accretion process itself, the convective heating effect on the protoplanet's upper atmospheric clouds is usually ignored.

In Fig. C1 we show a cartoon of a toy model where infalling (free-falling) hydrogen gas (thin blue arrows) impacts on the upper atmosphere of a protoplanet. The heating in the Zel'dovich shock temperature spike ($\Delta T_{shock}$ ~8000K) would reasonably drive strong convection into the upper atmosphere of the planet. The buoyancy from such a $\Delta T_{shock}$ would not be too dissimilar to that of the Shoemaker-Levy 9 impactors on Jupiter (H. Hammel et al. 1995). Using those observed convective plume heights (~5% of $R_{jup}$) as a guide, it is possible that "steady-state" cloud plumes of ~1-10% of the planet radius ($R_p$ ~1.6 $R_{jup}$) could occur. Absorption from these opacity sources (especially true of these dusty atmospheres where dust is well mixed in the upper atmosphere; I. Baraffe et al. 2002) along our line of slight to the H$\alpha$ shock can explain the puzzling lack of H$\alpha$ planets found with inclinations <37 degrees or >52 degrees (see Fig. 5). It also helps explains how, on the other hand, that ~80% of ALMA disks that have $37 \leq i \leq 52°$ and gap widths > 30 au are detected to have H$\alpha$ protoplanets (like WISPIT 2b) because at these angles we are looking straight down the accretion flow –with a direct line of slight to the Hα line formation region on the protoplanet (Fig. C1).

We caution that this is just a concept, other concepts are also possible and certainly may prove to be more likely. This toy model is presented here to start the discussion about inclination. Moroever, this model requires a rigorous theoretical modeling effort that properly accounts for the accretion heating, cloud formation/growth and radiative transfer through the opacity sources –but that is beyond the scope of this letter.

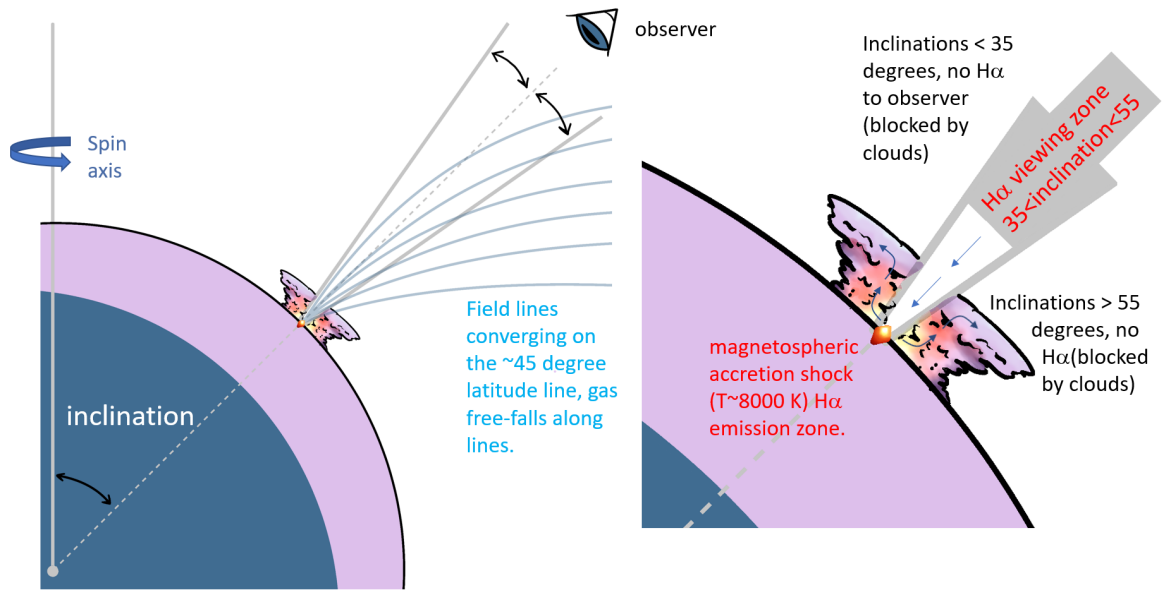

**Fig C1:** A simple cartoon to show how the Hα line formation zone from magnetospheric accretion could be best viewed by looking straight down the magnetic field lines (G. Marleau et al. 2022). In other words, an observer would have the most direct line of sight to the Hα line formation zone if the ALMA disk and its co-planar planet were both inclined at the line of latitude of where the magnetic field lines (drawn here fixed at 45º). Here we show a toy model where the heating in the Zel'dovich shock temperature spike ($\Delta T_{shock}$) would reasonably drive strong convection in the dusty atmosphere of the planet. The buoyancy from such a $\Delta T_{shock}$ would not be dissimilar to that of the Shoemaker-Levy 9 impactors on Jupiter and lead to the quasi-static ~5% $R_p$ dusty cloud deck heights shown here.